\documentclass[a4paper,10pt]{article}
\usepackage[utf8]{inputenc}
\usepackage[russian]{babel}
\usepackage{amssymb}\usepackage{amsmath}
\usepackage{graphicx, graphics}
\usepackage{float}

\def\be{\begin{equation}}
\def\ee{\end{equation}}
\newtheorem{dfn}{\bf Определение}

\def\ord{{\rm ord}}

\usepackage[usenames, dvipsnames, svgnames, table]{xcolor}
\usepackage[unicode]{hyperref}
\hypersetup{
     colorlinks   = true,
     allcolors    = NavyBlue 
}

\newcommand{\plot}[1]{
	\centerline{\includegraphics[width=\textwidth]{./#1.pdf}}
}

\begin{document}

\title{Аналитический подход к синтезу многополосных фильтров и его сравнение с другими подходами} 
\author{
	\copyright 2016 ~~~~ Богатырев А.Б., Горейнов С.А., Лямаев С.Ю.
	\thanks{Поддержано грантом РНФ 16-11-10349}
} 
\date{} 
\maketitle

В настоящее время многополосные фильтры широко применяются в аналоговой и цифровой технике. Например, в современной аппаратуре связи, работающей в СВЧ и радио- диапазонах, часто требуются фильтры с несколькими полосами пропускания, соотвествующими различным стандартам беспроводных коммуникаций (таким, как IEEE 806.11, IEEE 806.16, GSM, LTE, GPS, CDMA, DVBT/T2 и др.). Разработка фильтров с многополосными спецификациями представляет собой сложную инженерную проблему, особенно если требуется одновременно обеспечить и высокую производительность компонента, и его компактную реализацию, что и представляет интерес на практике. К текущему моменту в литературе не появилось полностью удовлетворительных и универсальных решений этой проблемы. 

Очевидным подходом является модульный: многополосный фильтр получается в результате соединения нескольких однополосных. Недостаток его -- существенная неоптимальность порядков синтезируемых фильтров, а значит и их массогабаритных характеристик. Этим объясняется интерес к разработке других подходов, идейно более сложных, но позволяющих получать фильтры, имеющие меньшие порядки при той же спецификации. Большую часть таких подходов можно условно разделить на следующие категории: 1) методы синтеза СВЧ фильтров, основанные на использовании многочастотных резонаторов; 2) методы, основанные на использовании частотных преобразований (например, \cite{2}, \cite{11}); 3) оптимизационные методы, основанные на алгоритме Ремеза (\cite{5}, \cite{7}, \cite{8} и др.).

Подходы первого типа предполагают синтез фильтра на основе мультимодальных резонаторов и позволяют достигать высокой степени миниатюрности синтезируемых устройств. При таком решении в формировании полос пропускания и задержки фильтра одновременно участвуют резонансы нескольких мод колебаний от каждого резонатора, за счет чего уменьшается число резонаторов, а потому и массо-габаритные характеристики фильтра. Однако подходы этого типа специфичны только для СВЧ устройств, причем дизайн и реализация фильтров высокого порядка с их помощью представляет существенную трудность. Методы второго типа – синтез фильтра на основе многополосного частотного преобразования – являются аналитическими либо полуаналитическими, и осуществляют приведение однополосного (широкополосного) фильтра-прототипа к многополосной конфигурации за счет использования замен частотной переменной. Сложность таких преобразований быстро растет с увеличением числа полос пропускания, а обслуживаемые спецификации, как правило, ограничиваются симметричными.

Большое внимание в технической литературе уделяется развитию методов третьего типа, связанных с синтезом оптимальных фильтров и близких к ним. Под оптимальным для заданной многополосной спецификации в данном случае понимается фильтр, который обладает наименьшим порядком среди всех физически осуществимых фильтров, удовлетворяющих требованиям этой спецификации. Скажем также, что под спецификацией (маской) фильтра здесь имеется в виду совокупность следующих параметров: 1) значения желаемых граничных частот всех полос пропускания и задержки; 2) максимальная допустимая величина неравномерности АЧХ на полосах пропускания и минимальная допустимая величина подавления в полосах задержки. Эти данные определяют желаемый вид амплитудно-частотной характеристики фильтра, не затрагивая других его характеристик (фазо-частотной, импульсной и т.д.). 

Все существующие методы синтеза оптимальных многополосных фильтров основываются на прямой численной оптимизации с помощью алгоритмов типа Ремеза, и, ввиду присущей для такой оптимизации принципиальной неустойчивости, имеют ограниченную область применимости: с их помощью возможен синтез фильтров с числом полос пропускания, как правило, не превышающим трех, и порядками, как правило, не выше 20. 

Авторами разработан аналитический подход к синтезу оптимальных многополосных фильтров, который может быть применен для спецификаций с общим числом полос (на текущий момент) вплоть до 23 и порядками вплоть до 1000. Целью работы является сравнение нового подхода к синтезу оптимальных многополосных фильтров с методом прямой оптимизации, основанном на алгоритме типа Ремеза, а также сравнение оптимальных фильтров с неоптимальными составными фильтрам, получаемыми модульным подходом. В первом разделе ставится оптимизационная задача, лежащая в основании проблемы синтеза оптимального многополосного фильтра. Второй раздел содержит краткое описание нового аналитического подхода к решению этой задачи. В третьем разделе приводятся рецепты, использованные нами при решении той же задачи методом прямой оптимизации. В четвертом разделе описывается модульный подход к синтезу неоптимальных многополосных фильтров. Пятый раздел содержит результаты численных экспериментов, посвященных сравнению этих трех подходов. В приложении приводятся графики фазо-частотной и импульсной характеристик оптимального четырехполосного фильтра.

Авторы благодарны проф. Л. Баратшару за обсуждение задач равномерного рационального приближения и д-рам Ф. Зайферту и В. Люно (INRIA, Sophia-Antipolis) за предоставленный код, позволяющий рассчитывать оптимальные фильтры алгоритмом типа Ремеза.  

\section{Постановка задачи}
	Поиск оптимального фильтра по заданной спецификации может быть сведен к решению серии задач о наименьшем уклонении типа сформулированных ниже. Для этого на каждом шаге фиксируется порядок фильтра и максимизируется величина подавления в полосах задержки при фиксированных остальных параметрах спецификации. Приведем две встречающиеся в литературе формулировки.

	Пусть $E$ --- множество, состоящее из $ m $ непересекающихся отрезков действительной оси (частотных диапазонов), разделенное на два подмножества: полосы пропускания $ E_+ $ и полосы задержки $ E_- $. Скажем, что идеальная функция пропускания $ F $ равна $ 1 $ на отрезках из $ E_+ $ и $ -1 $ на отрезках из $ E_- $. 

	{\bf Задача 1}.{\it Найти вещественную рациональную функцию $ R_n $ степени не выше $ n $, для которой величина отклонения от функции пропускания $ F $ минимальна в равномерной норме на $ E $:}
	\begin{equation}
		||R_n-F||_{C(E)} := \max\limits_{x \in E}|R_n(x) - F(x)| \to \min =: \mu.
		\label{eq:zol_prob}
	\end{equation}

	{\bf Задача 2}. {\it Найти вещественную рациональную функцию $Q_n(x)$ степени не выше $n$, минимизирующую величину $ \theta $ при условиях}
	\begin{equation}
		\min_{w\in E_-}|Q_n(w)|\ge\theta^{-1},\quad \max_{w\in E_+}|Q_n(w)|\le\theta.
		\label{eq:mmx_prob}
	\end{equation}

Нетрудно показать, что эти задачи эквивалентны, и их решения отличаются дробно-линейной подстановкой, то есть $ Q_n=f\circ R_n$, где $ f $ -- дробно-линейное преобразование, а обратные к минимальным отклонениям связаны преобразованием Жуковского $ \mu^{-1}=(\theta+\theta^{-1})/2 $.

Отметим, что сформулированная задача о рациональной аппроксимации есть третья задача Золотарева для конденсатора $ (E_+, E_-) $ \cite{Gonchar}. Она является многоэкстремальной: все множество рациональных функций разбивается на $2^{m-2}$ непересекающихся класса \cite{9}, \cite{Malo}, в каждом из которых решение существует, единственно и допускает альтернансную характеризацию: на $E$ имеется $2n+2$ точек альтернанса, в которых для решения степени $ n $ достигается величина отклонения с последовательной переменой знака \cite{Akh}. 

До недавнего времени аналитическое решение этой задачи было известно только для случая $m=2$ (для одной полосы пропускания и одной полосы задержки), найденное в 1870-х годах учеником П.Л. Чебышёва Е.И. Золотарёвым \cite{Zol}. Именно оно около полувека спустя было использовано немецким учёным-электротехником В. Кауэром \cite{Cauer} для расчёта передаточных функций оптимальных фильтров низких частот, получивших название фильтров Кауэра-Золотарёва (эллиптических). К настоящему времени эти фильтры нашли широкое применение в аналоговой и цифровой технике, а метод аппроксимации АЧХ по Золотарёву-Кауэру стал классическим.

\section{Новый аналитический подход}

Подход к решению задач \eqref{eq:zol_prob}, \eqref{eq:mmx_prob}, предлагаемый в данной статье, опирается на явные аналитические формулы, полученные в \cite{1} и обобщающие решение Золотарева на случай $ m > 2 $. Идея этого подхода следующая: как было отмечено выше, все решения степени $n$ задачи о наименьшем уклонении имеют $2n+2$ точки альтернанса на $ E $, причем каждая из них, находящаяся во внутренности множества $ E $, с необходимостью будет критической точкой функции-решения $ R(x) $ со значением во множестве из четырех элементов $ \pm1 \pm \mu $ для формулировки \eqref{eq:zol_prob} и $ \pm \theta,~~\pm \theta^{-1} $ для формулировки \eqref{eq:mmx_prob}. Всего же рациональная функция степени $ n $ имеет с учетом кратности $2n-2$ критических точки, поэтому решения рассматриваемых задач о наименьшем уклонении удовлетворяют следующему определению с малым параметром $g$:

\begin{dfn}
Рациональная функция $R(x)$ называется $g-$экстремальной относительно 
4-элементного множества значений ${\sf Q}$, если все ее критические точки -- за исключением $g$ из них -- простые со значениями в $\sf Q$.
Число исключительных критических точек подсчитывается по формуле
\begin{equation}
\label{g}
g=1+ \sum\limits_{x: \,
R(x)\not\in{\sf Q}} \operatorname{ord} \,  dR(x) \, + \, 
\sum\limits_{x: 
\, R(x)\in{\sf Q}}
\left[
\frac12\ord \,  dR(x)
\right],
\end{equation}
где суммирование производится по всем точкам сферы Римана; $\ord \; dR(x)$ -- порядок нуля 
дифференциала голоморфного отображения $R: {\mathbb C}P^1\to {\mathbb C}P^1$ в
точке $x$ (например, в простых полюсах функции $R(x)$ это число равно нулю), и $ [\cdot] $ -- целая часть числа.
\end{dfn} 

Функции с малым числом экстремальности весьма специфичны и заполняют многообразия малой размерности в пространстве всех рациональных функций. При решении задач о наименьшем уклонении возможно перейти от поиска по всему пространству рациональных функций к 
поиску по этим многообразиям. Экстремальные рациональные функции обладают следующим 
эффективным малопараметрическим представлением \cite{1}:
\begin{equation}
\label{R} 
R(x)=
\mathop{\rm sn} 
\left(\int_{(e,0)}^{(x,w)}d\zeta + A(e)\,\biggl|\,\tau\right),
\qquad  {\sf Q}=\{\pm1, \pm1/k(\tau)\},
\end{equation}
где $d\zeta$ -- голоморфный дифференциал на римановой поверхности рода $g$, периоды которого лежат в решетке периодов эллиптического синуса, и фазовый сдвиг $A(e)$ также соизмерим с этой решеткой. Эта поверхность определяется по рациональной функуции $R(x)$ как двулистная накрывающая сферы Римана с ветвлением в точках, где $R(x)$  принимает значения из $\sf Q$ с нечётной кратностью. 
Возникаюшая  риманова поверхнось не произвольна, это так называема кривая Калоджеро-Мозера:
она разветвленно накрывает тор, определяемый по множеству выделенных значений $\sf Q$.

Используемая параметризация экстремальных функций позволяет контролировать поведение решения в переходных полосах фильтра (выбор класса) и непосредственно решать задачу о фильтре наименьшей степени для заданной спецификации без рассмотрения цепочки задач о наименьшем уклонении. Вычислительные средства, используемые для нахождения экстремальных рациональных функций по явной аналитической формуле, включают разработанный ранее \cite{1, B, B2, B3} аппарат эффективных вычислений на римановых поверхностях, и позволяют устойчиво вычислять решения степени  $n$ равной по меньшей мере нескольким сотням. 

\begin{figure}[H]{\plot{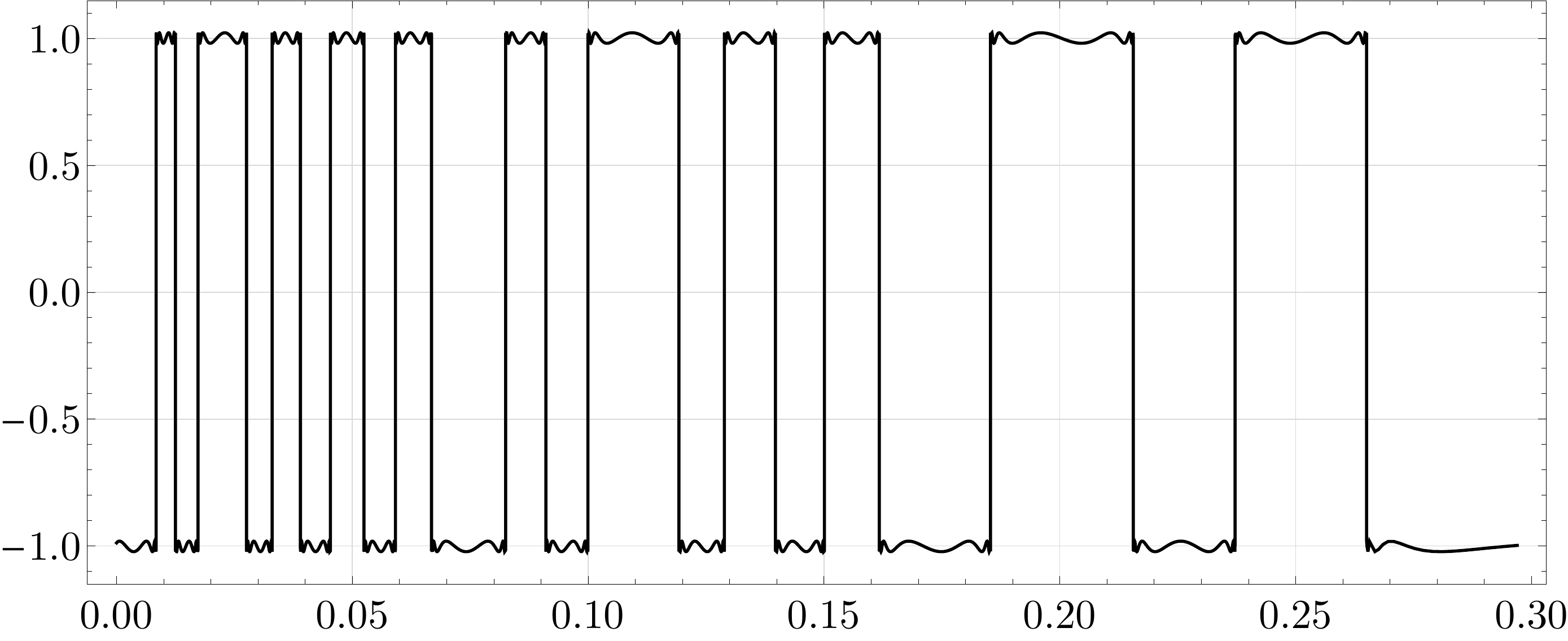}\caption{График решения оптимизационной задачи (для спецификации с $ m = 23 $)}\label{ex1}}\end{figure}

\begin{figure}[H]{\plot{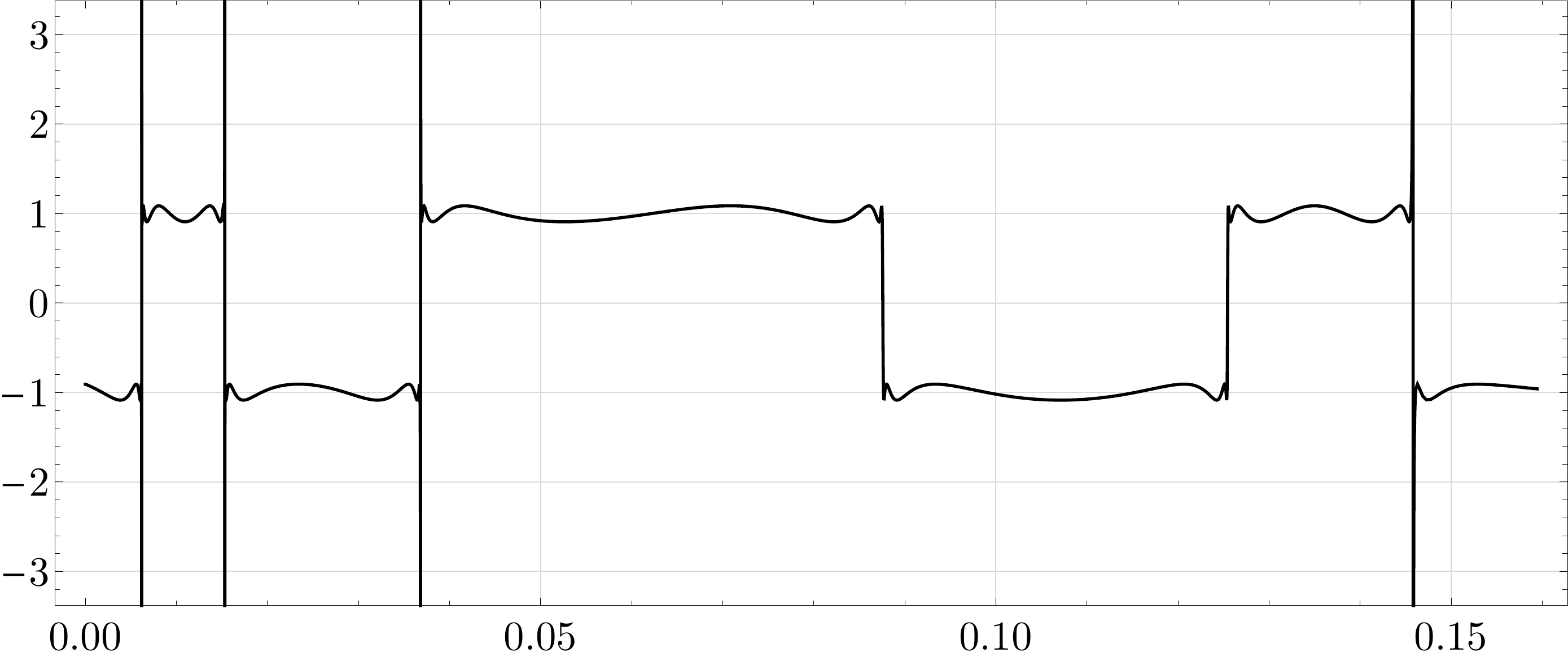}\caption{График решения оптимизационной задачи (для спецификации с $ m=7 $)}\label{ex2}}\end{figure}

С использованием нового аналитического подхода были рассчитаны примеры решений оптимизационной задачи \ref{eq:zol_prob}. На Рис. \ref{ex1} представлен график функции-решения степени 654 для спецификации с числом отрезков $ m=23 $. На Рис. \ref{ex2} изображен график решения (для спецификации с $ m= 7$) из класса, допускающего наличие полюсов в первой, второй, третьей и шестой переходных полосах.

Детальное изложение алгоритма, основанного на использовании явной аналитической формулы \eqref{R} будет приведено в отдельной работе, поскольку требует сложного математического аппарата. 

\section{Прямая оптимизация}
Решения оптимизационной задачи могут быть вычислены приближенно, например, алгоритмами типа Ремеза \cite{Remez, Veidinger, 9}.
Отправной точкой снова служит теорема Ахиезера: оптимальная рациональная функция $\phi(w)/\psi(w)$ степени $n$ должна иметь на $E$ по крайней мере $2n+2$ точки альтернанса. Предположим, что мы имеем приближенную дробь, заданную,
например, положением нулей и полюсов или коэффициентами
числителя и знаменателя по некоторому полиномиальному базису;
и, кроме того, приближенные точки альтернанса, множество которых обозначим за $A$.
Рассматриваемые алгоритмы состоят в поочередном уточнении каждого из этих двух объектов:
\begin{enumerate}
\item имея множество точек $A\subset E$, можно изменить дробь так, чтобы на 
дискретном множестве $A$ соблюдался альтернанс;
\item имея дробь, можно изменить положение точек с целью увеличения на $A$ равномерной нормы ошибки приближения -- или же убедиться, что это невозможно, то есть равномерная норма ошибки не уменьшается при ограничении последней с $E$ на $A$.
\end{enumerate}
Кроме того, требуется начальное приближение, выбор которого из-за локальной сходимости
алгоритмов типа Ремеза весьма важен.
Изложим кратко использованные нами рецепты для всех указанных шагов.

{\it Выбор начального приближения}.
Мы начинаем с выбора множества $A$, именно, используются $1/(2n+2)$-квантили
равновесной меры для множества $E$, полученной при помощи численного решения
интегрального уравнения с логарифмическим ядром \cite{Fuchs}.

{\it Шаг 2 (уточнение $A$ по заданной дроби $\phi/\psi$)}.
Рассматривается сначала более широкое множество, составленное
из концов наших $m$ интервалов и критических точек функции ошибки $\phi(w)/\psi(w)-F(w)$, лежащих на $ E $.
Точки, в которых отклонение меньше полученного ранее на шаге~1,
отбрасываются.
Новое множество $A$ должно подчиняться правилу перемены знаков
и содержать $2n+2$ точки; наличие меньшего или большего их количества
исправляется с помощью введенной ранее равновесной меры;
например, выбрасываются точки, соответствующие меньшим значениям плотности этой меры.
Поиск критических точек осуществляется методом Брента или, если
приближение достаточно хорошее, проще --- методом Ньютона.

{\it Шаг 1 (уточнение дроби $\phi/\psi$ по заданному $A$)}.
В отличие от полиномиальных приближений, условие на альтернанс на множестве $A$,
по всей видимости, невозможно представить в виде линейной системы; в лучшем случае
возникают обобщенные задачи на собственные значения, вида
\be\label{eq:appr_evp}
\begin{pmatrix}V_+& -V_+\\
V_- &V_-\end{pmatrix}
\begin{pmatrix}\phi\\ \psi
\end{pmatrix}=\mu
\begin{pmatrix}0& \Sigma_+V_+\\
               0& \Sigma_-V_-
\end{pmatrix}
\begin{pmatrix}\phi\\ \psi
\end{pmatrix}
\ee
для первой формулировки экстремальной задачи, или вида
\be\label{eq:mmx_evp}
\begin{pmatrix}V_-& 0\\
0 &\Sigma_+V_+\end{pmatrix}
\begin{pmatrix}\phi\\ \psi
\end{pmatrix}=\theta
\begin{pmatrix}0& \Sigma_-V_-\\
               V_+& 0
\end{pmatrix}
\begin{pmatrix}\phi\\ \psi
\end{pmatrix}
\ee
для второй формулировки (см. раздел 1).
Последнюю задачу можно найти в \cite{9}.
В уравнениях (\ref{eq:appr_evp}--\ref{eq:mmx_evp}) использованы обозначения:
$V_\pm$ --- матрица Вандермонда по узлам $A$, принадлежащим $E_\pm$;
$\Sigma_\pm$ --- диагональные матрицы с элементами $ \pm 1 $, фиксированными
для каждого из интервалов пропускания или задержки,
по узлам $A$, принадлежащим $E_\pm$;
$\phi$, $\psi$ --- векторы коэффициентов числителя и знаменателя
рациональной дроби; $\mu$, $\theta$ --- экстремальные альтернирующие значения.
Наличие матриц $\Sigma_\pm$ связано с возможным изменением
знака альтернанса в соседних интервалах.

Как видно, матричные задачи 
(\ref{eq:appr_evp}--\ref{eq:mmx_evp}) несимметричны, и соответствующие
пучки могут оказаться сингулярными; 
даже в случае двух интервалов может оказаться, что
стандартной мантиссы не хватит для вычисления хотя бы одного знака коэффициентов 
$\phi$, $\psi$ (не говоря уже о корнях).
Кроме того, перебор всех вариантов знаков альтернанса в соседних интервалах
означает экспоненциальную по $m$ сложность.

Некоторым преимуществом, по сравнению с задачами на собственные значения,
обладает редукция задачи \eqref{eq:zol_prob}, сформулированной
относительно заданного множества $A$, к задаче линейного программирования,
причем и здесь возникает экспоненциальный по $m$ перебор, связанный с выбором знака
знаменателя дроби на каждом интервале. 
Вот эта задача: потребуем минимизации величины $t\in\mathbb{R}$
при ограничениях
$$
-t\sigma(w)\psi(w)\le\sigma(w)\phi(w)-\sigma(w)\psi(w)F(w)\le t\sigma(w)\psi(w),\qquad 
w\in A,\quad j=1,\dots,m,
$$
причем мы предполагаем ${\rm sign}\psi(w)=:\sigma(w)$ априори заданным на каждом интервале.
Формально от последнего предположения можно освободиться некоторым увеличением числа переменных.
Хорошие результаты для такой задачи показывает метод внутренней точки
для объединенных прямой и двойственной формулировок \cite{Boyd}.

Линейная параметризация числителей и знаменателей дроби существенно ограничивает устойчивость 
алгоритмов типа Ремеза \cite{Remez}. Максимальная степень решения $R$, полученная таким методом при 
вычислениях с двойной точностью зависит от конфигурации множества $E$ и не превосходит $n=20$. 

\section{Модульный подход}
	
	Модульный подход не связан с решением сформулированной оптимизационной задачи и предполагает получение (неоптимального) многополосного фильтра в результате соединения нескольких однополосных. Архитектурно это может осуществляться, например, путем параллельного включения полосно-пропускающих фильтров, каждый из которых обслуживает одну рабочую полосу из заданной многополосной спецификации, либо путем каскадного соединения полосно-пропускающих и полосно-заграждающих фильтров. 
	
	Неоптимальные фильтры, получаемые модульным подходом, ниже мы называем составными. Нами использовался следующий рецепт их построения: по каждой полосе пропускания рассчитывалась передаточная функция соотвествующего полосового эллиптического фильтра, затем полученные функции складывались. Параметры эллиптических фильтров оптимизировались ручным перебором для достижения как можно меньшего порядка итогового многополосного фильтра при условии, чтобы его АЧХ была вписана в заданный спецификацией коридор.

\section{Примеры синтеза}

	В этом разделе приводятся примеры синтеза цифровых многополосных фильтров с помощью трех подходов: нового аналитического подхода, метода прямой оптимизации и модульного подхода. Первые два подхода, основываются на решении сформулированной выше оптимизационной задачи, и дают оптимальные фильтры (то есть порядка, минимального достижимого для заданной спецификации), последний подход заключается в соединении однополосных эллиптических фильтров и дает неоптимальные многополосные фильтры.
	
	Оптимальные цифровые фильтры получались из оптимальных аналоговых путем стандартного, так называемого билинейного, преобразования частоты. 
	
	Прямая оптимизация ни для одного из рассмотренных примеров не дала удовлетворительных результатов, ввиду сложности спецификаций. Её результаты приведены для первого и второго примера. Модульный подход, как видно из примеров, дает фильтры с существенно большими порядками, в сравнении с оптимальными.
	
	\subsection{Однополосный фильтр} 
\begin{figure}[H]{\plot{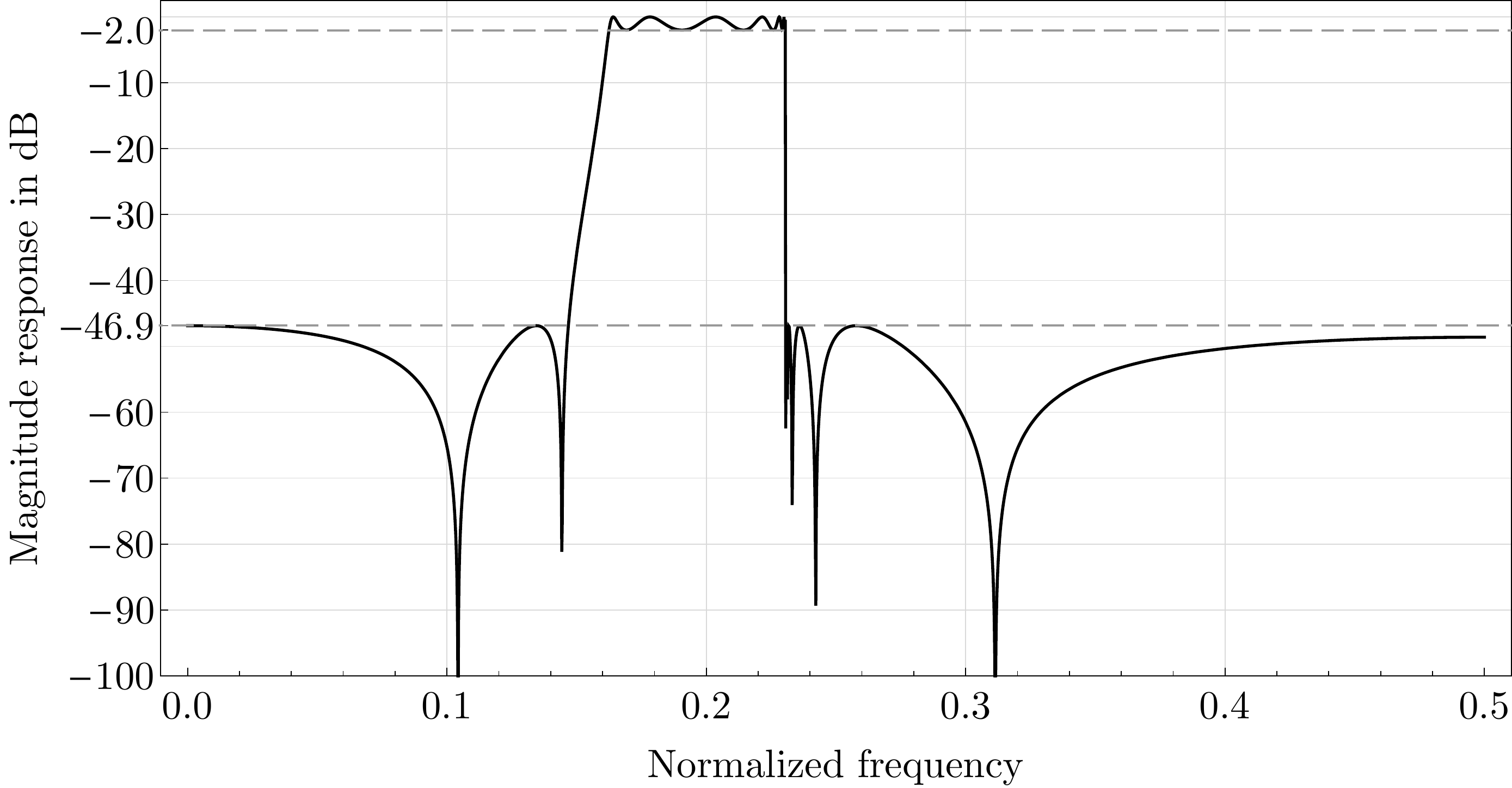}\caption{АЧХ однополосного оптимального фильтра}\label{1b}}\end{figure}

		С помощью нового аналитического подхода нами был рассчитан однополосный оптимальный фильтр 18-ого порядка с сильно несимметричными ширинам переходных полос, равными 0.016 и $ 2 \cdot 10^{-5} $. График его АЧХ приводится на Рис. \ref{1b}.
		
\begin{figure}[H]{\plot{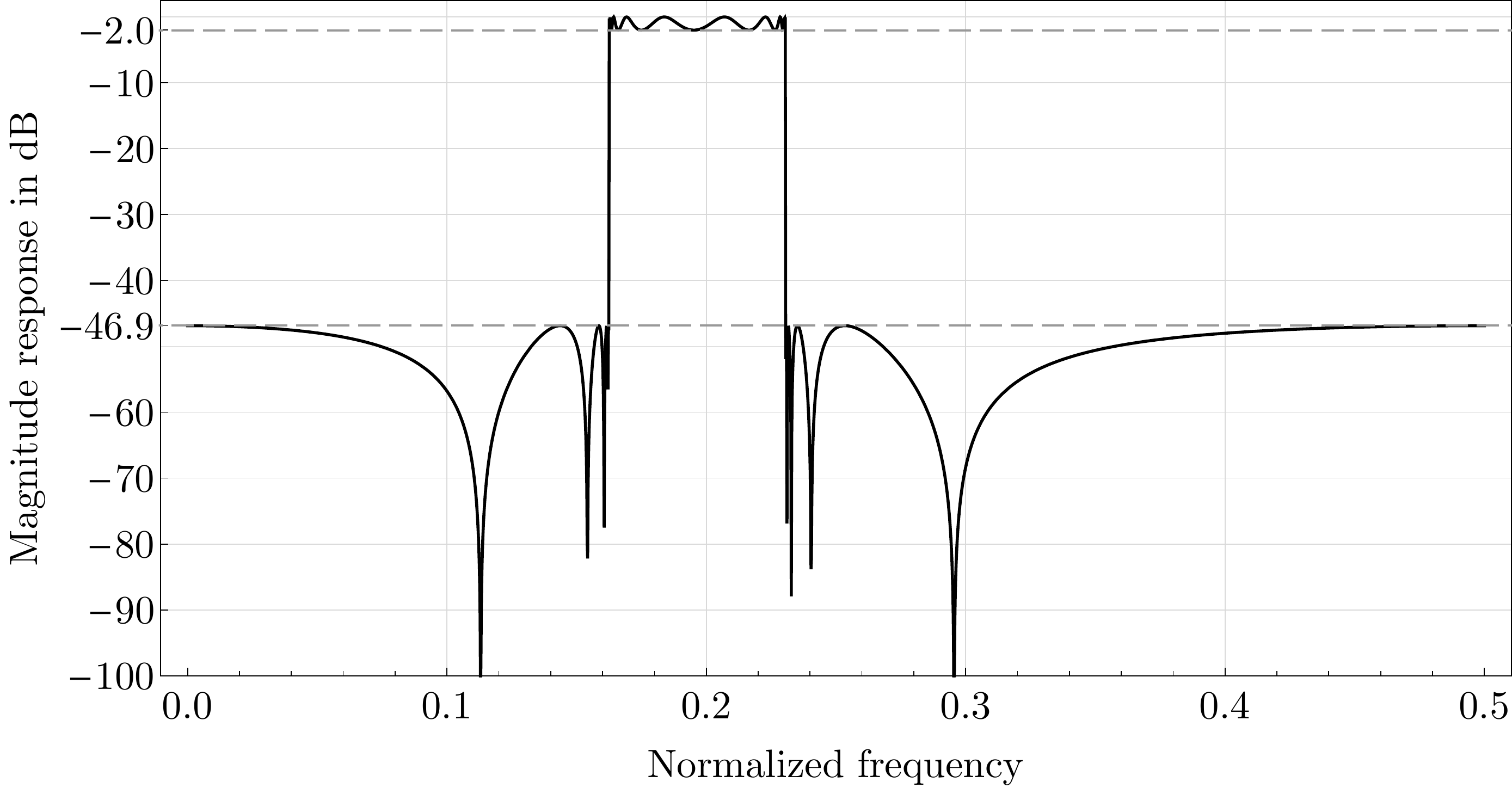}\caption{АЧХ однополосного эллиптического фильтра}\label{1be}}\end{figure}

		Стандартное частотное преобразование, используемое при синтезе полосового эллиптического фильтра из НЧ-прототипа, не может обеспечить такой разницы в ширинах переходных полос, и эллиптический фильтр, рассчитанный по той же спецификации, имеет больший порядок -- 28. График его АЧХ представлен на Рис. \ref{1be}
		
		Результаты прямой численной оптимизации для такой спецификации приведены в Табл. \ref{tb:do_one}. При расчетах с помощью алгоритма типа Ремеза границы полос фильтра задавались спецификацией, неравномерность в полосе пропускания фиксировалась равной 2 dB, а затухание в полосах задержки (второй столбец в таблице) определялось, исходя из порядка получаемого оптимального фильтра. Затухание в полосах задержки в $ -46.9 $ dB для метода прямой оптимизации оказалось недостижимым ввиду того, что уже для расчета фильтра 10 порядка требовались большие временные затраты.
	
		\begin{table}[H]
		\label{tb:do_one}
		\begin{center}
		\begin{tabular}{rcc}
			Порядок & Затухание в полосе задержки & Число итераций\\ 
			\hline
			3       & -3.3812                     & 102\\
			4       & -4.2272                     & 154\\
			5       & -7.5894                     & 211\\
			6       & -10.612                     & 308\\
			8       & -19.355                     & 457\\
			9       & -22.378                     & 652\\
			10      & ---                         & $>$1000\\ 
			\hline\hline\\[-2mm]
			\bf18   & \bf-46.9                    &	\bf?\\
		\end{tabular}
		\caption{Результаты прямой численной оптимизации для однополосного фильтра.}
		\end{center}
		\end{table}

	\subsection{Двухполосный фильтр} 
\begin{figure}[H]{\plot{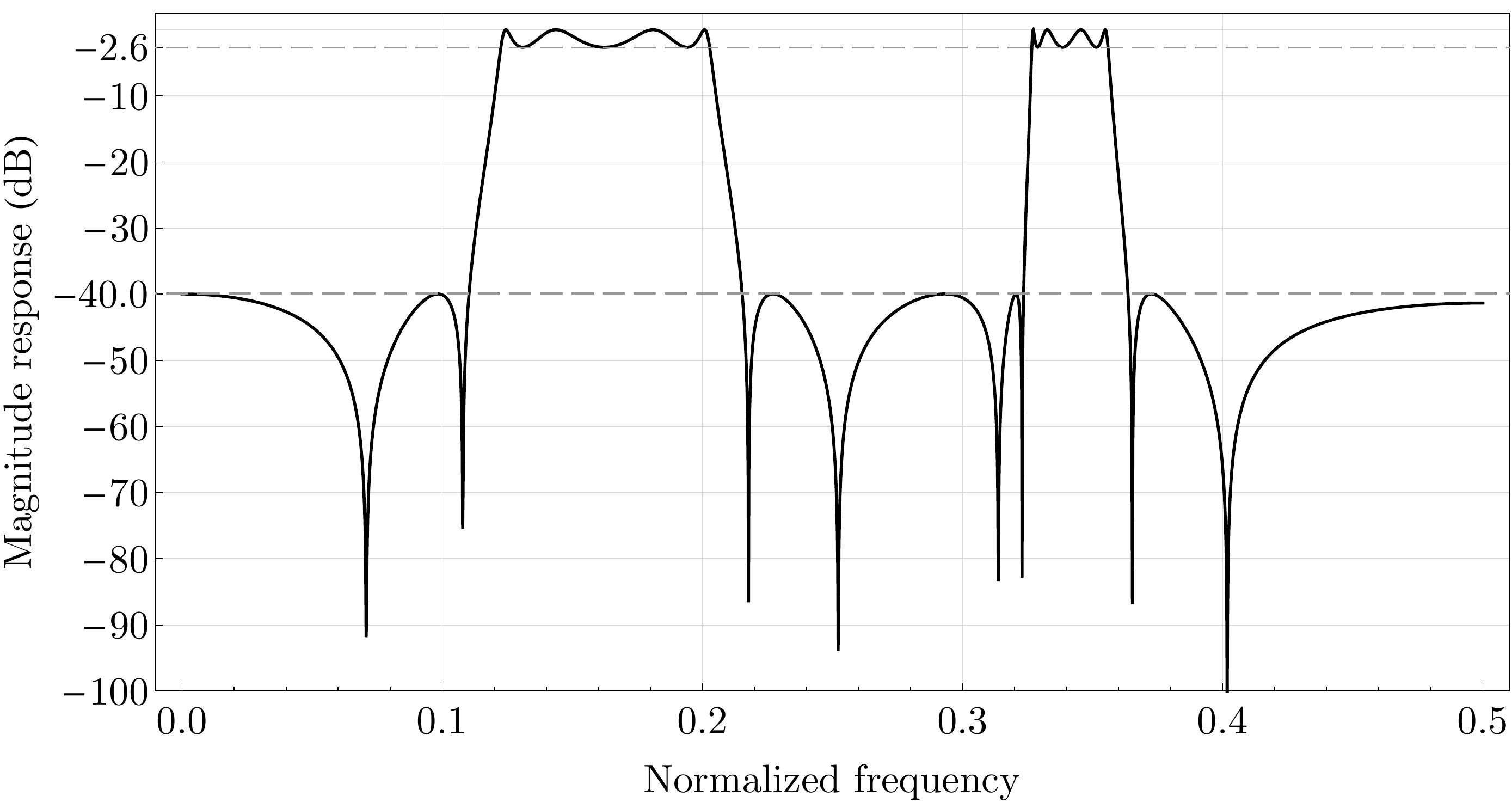}\caption{АЧХ двухполосного оптимального фильтра}\label{2b}}\end{figure}

		В качестве примера оптимального двухполосного фильтра с помощью аналитического подхода был рассчитан фильтр 16-ого порядка с минимальным затуханием в полосах задержки в 40 dB, и величиной пульсаций АЧХ в полосах пропускания в 2.6 dB. Ширины переходных полос: 0.012, 0.012, 0.003, 0.008. График АЧХ приведен на Рис. \ref{2b}.
	
\begin{figure}[H]{\plot{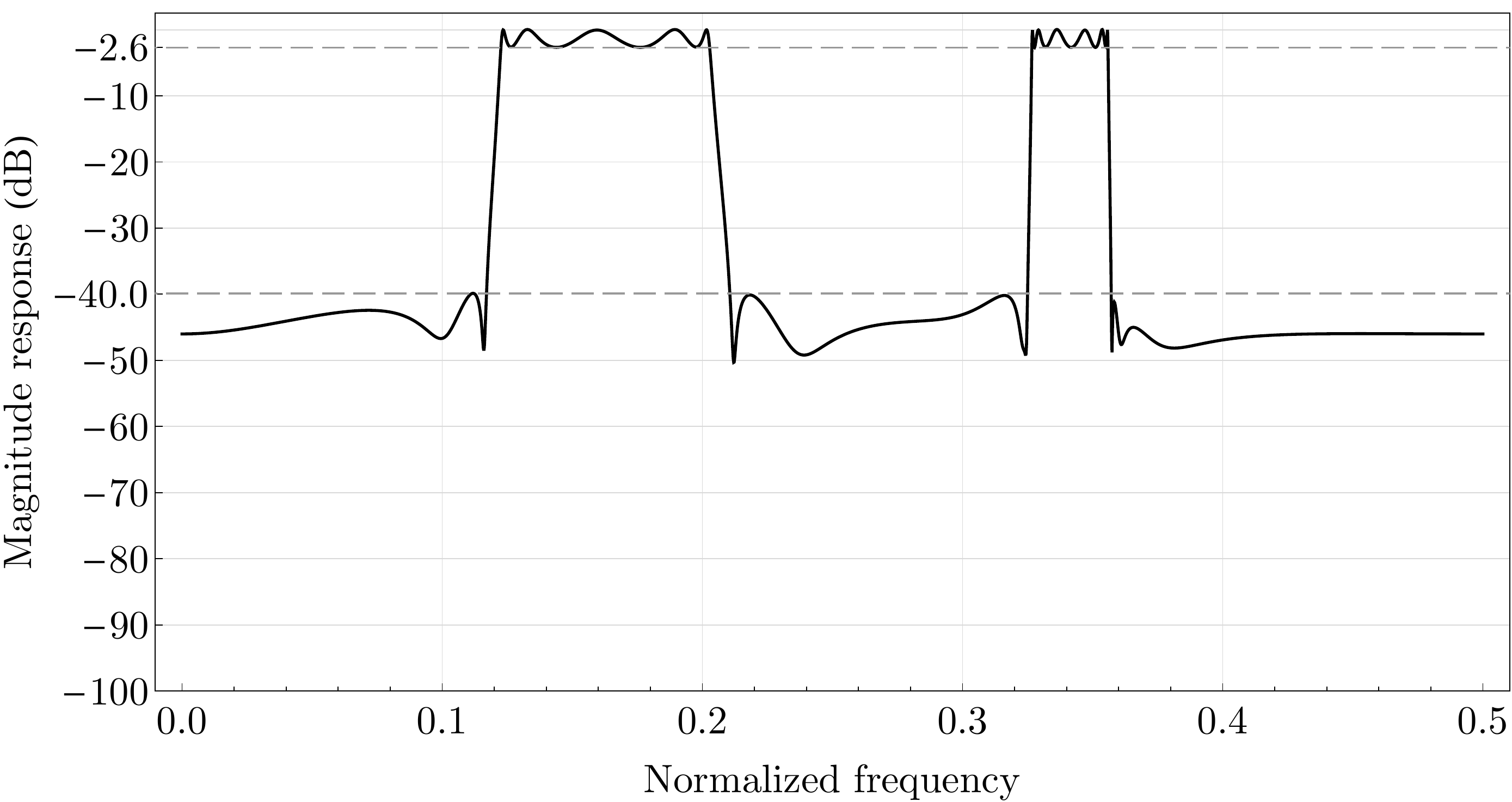}\caption{АЧХ двухполосного составного фильтра}\label{2be}}\end{figure}

		Построенный по той же спецификации составной фильтр имеет порядок 23. График его АЧХ представлен на Рис. \ref{2be}.
			
		Результаты прямой оптимизации для такой спецификации представлены в Табл. \ref{tb:do_two}. При расчетах границы полос фильтра задавались спецификацией, неравномерность в полосе пропускания фиксировалась равной 2 dB, а затухание в полосах задержки (второй столбец в таблице) определялось, исходя из порядка получаемого оптимального фильтра. В данном случае алгоритм типа Ремеза оказался неприменимым для поиска оптимальных фильтров начиная уже с восьмого порядка, соответственно, оптимальный фильтр 16-ого порядка, найденный с помощью аналитического подхода, метод прямой оптимизицации рассчитать не позволяет.
		\begin{table}[H]
		\begin{center}
		\begin{tabular}{rcc}
		Порядок & Затухание в полосах задержки & Число итераций\\ 
		\hline
		3       & -5.3421                     & 151\\
		4       & -9.1785                     & 273\\
		5       & -12.649                     & 351\\
		6       & -16.013                     & 408\\
		7       & -19.215                     & 594\\
		8       & ---                         & $>1000$\\ 
		\hline\hline\\[-2mm]
		\bf16   & \bf-40                      & \bf?\\
		\end{tabular}
		\caption{Результаты прямой оптимизации для двухполосного фильтра.}
		\label{tb:do_two}
		\end{center}
		\end{table}
	
	\subsection{Четырёхполосный фильтр}\label{4bsection}
\begin{figure}[H]{\plot{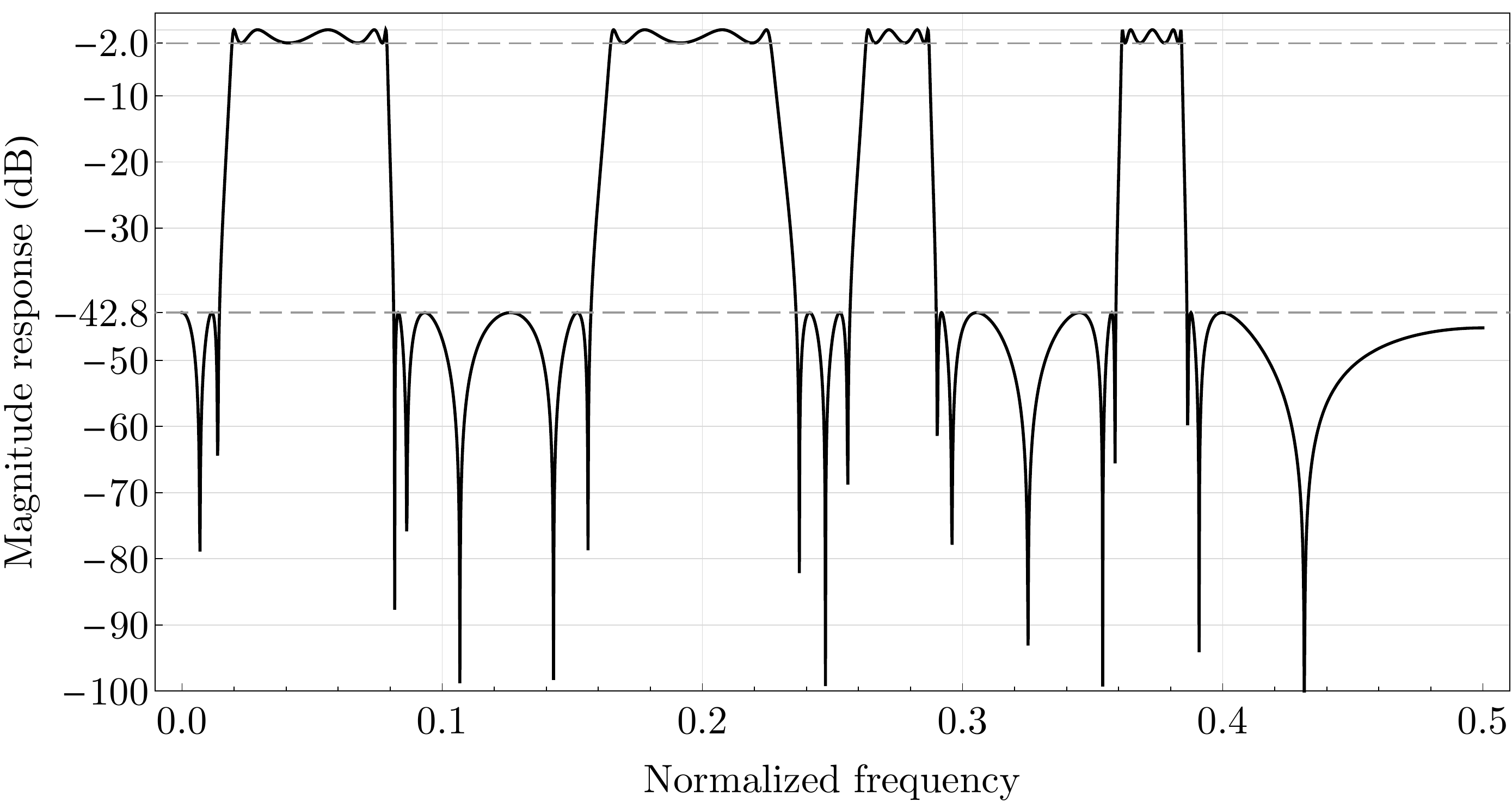}\caption{АЧХ четырёхполосного оптимального фильтра}\label{4b}}\end{figure}

		С помощью аналитического подхода нами был построен четырёхполосный оптимальный фильтр 36-ого порядка с минимальным затуханием в полосах задержки в 42.8 dB, неравномерностью АЧХ в полосах пропускания в 2.0 dB, и ширинами переходных полос от 0.002 до 0.005. График АЧХ полученного фильтра приведен на Рис. \ref{4b}.
		
\begin{figure}[H]{\plot{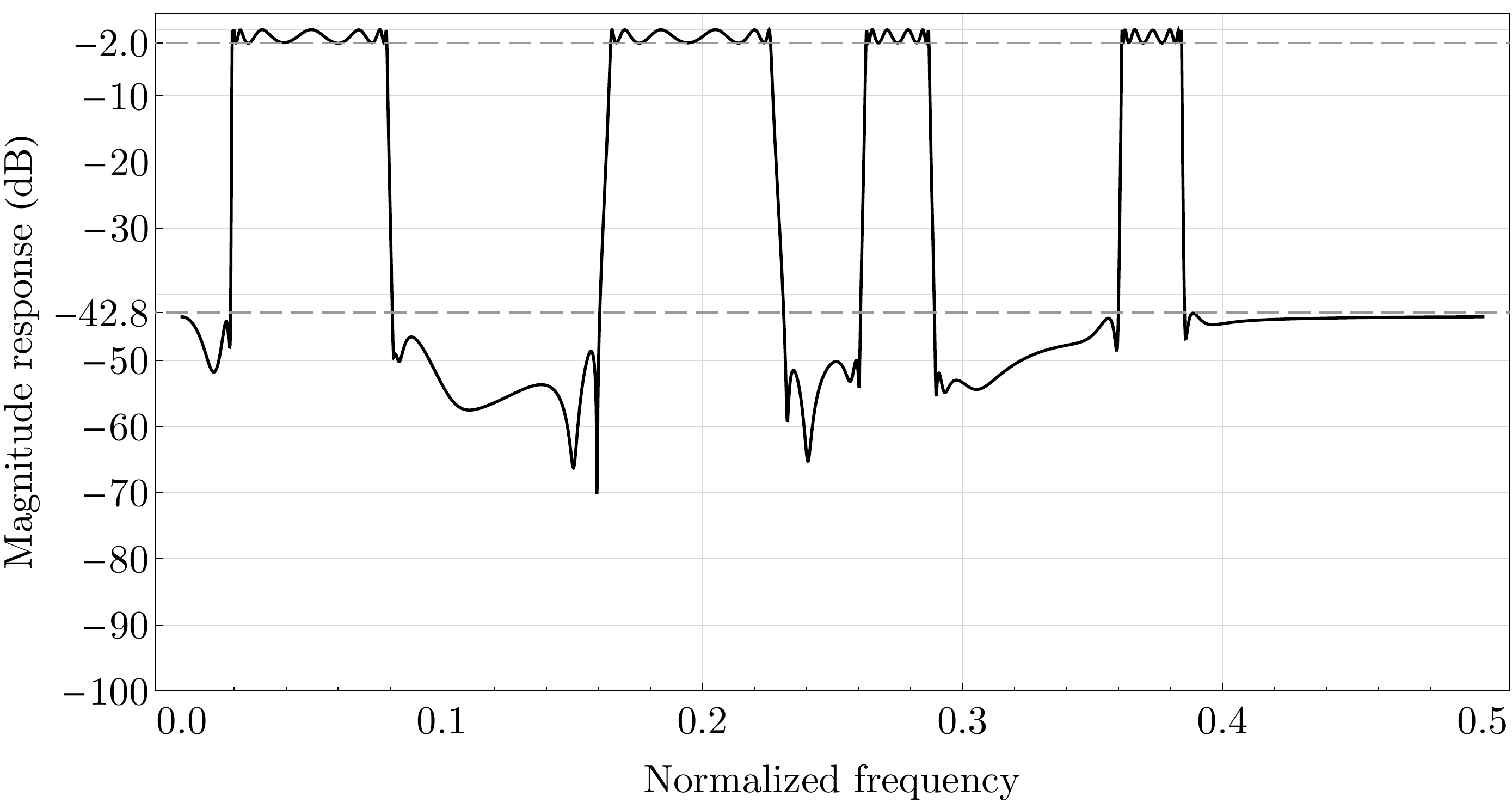}\caption{АЧХ четырёхполосного составного фильтра}\label{4be}}\end{figure}

		Рассчитанный по той же спецификации составной фильтр имеет порядок 55. Его АЧХ приводится на Рис. \ref{4be}.
	
	\subsection{Пятиполосный фильтр}
\begin{figure}[H]{\plot{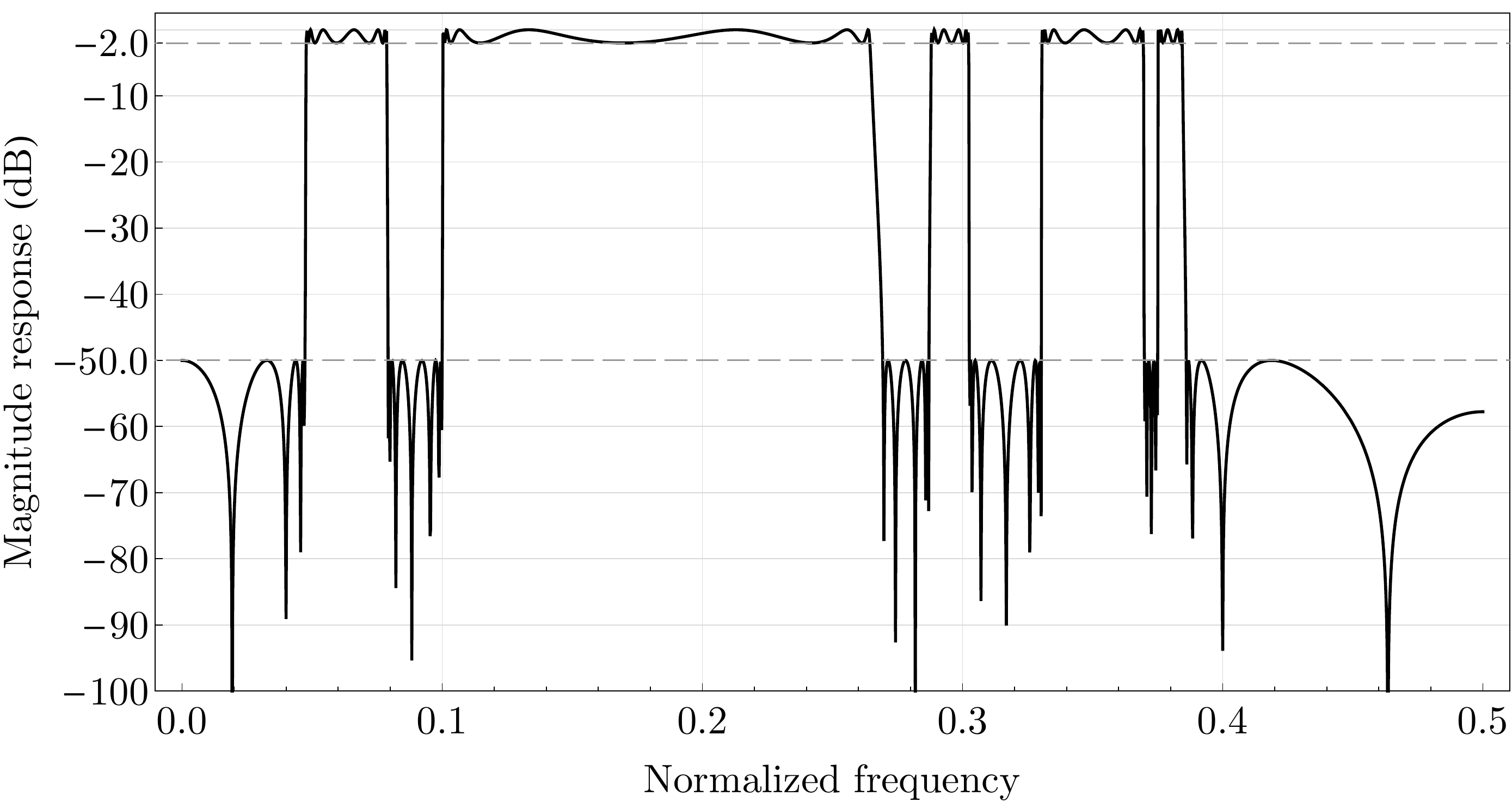}\caption{АЧХ четырёхполосного оптимального фильтра}\label{5b}}\end{figure}

		Для пятиполосной спецификации был рассчитан оптимальный фильтр 76-ого порядка с минимальным затуханием АЧХ в полосах задержки в 50 dB, и неравномерностью в полосах пропускания в 2.0 dB. Ширины переходных полос лежат в промежутке от 0.002 до 0.005. График АЧХ представлен на Рис. \ref{5b}.
		
\begin{figure}[H]{\plot{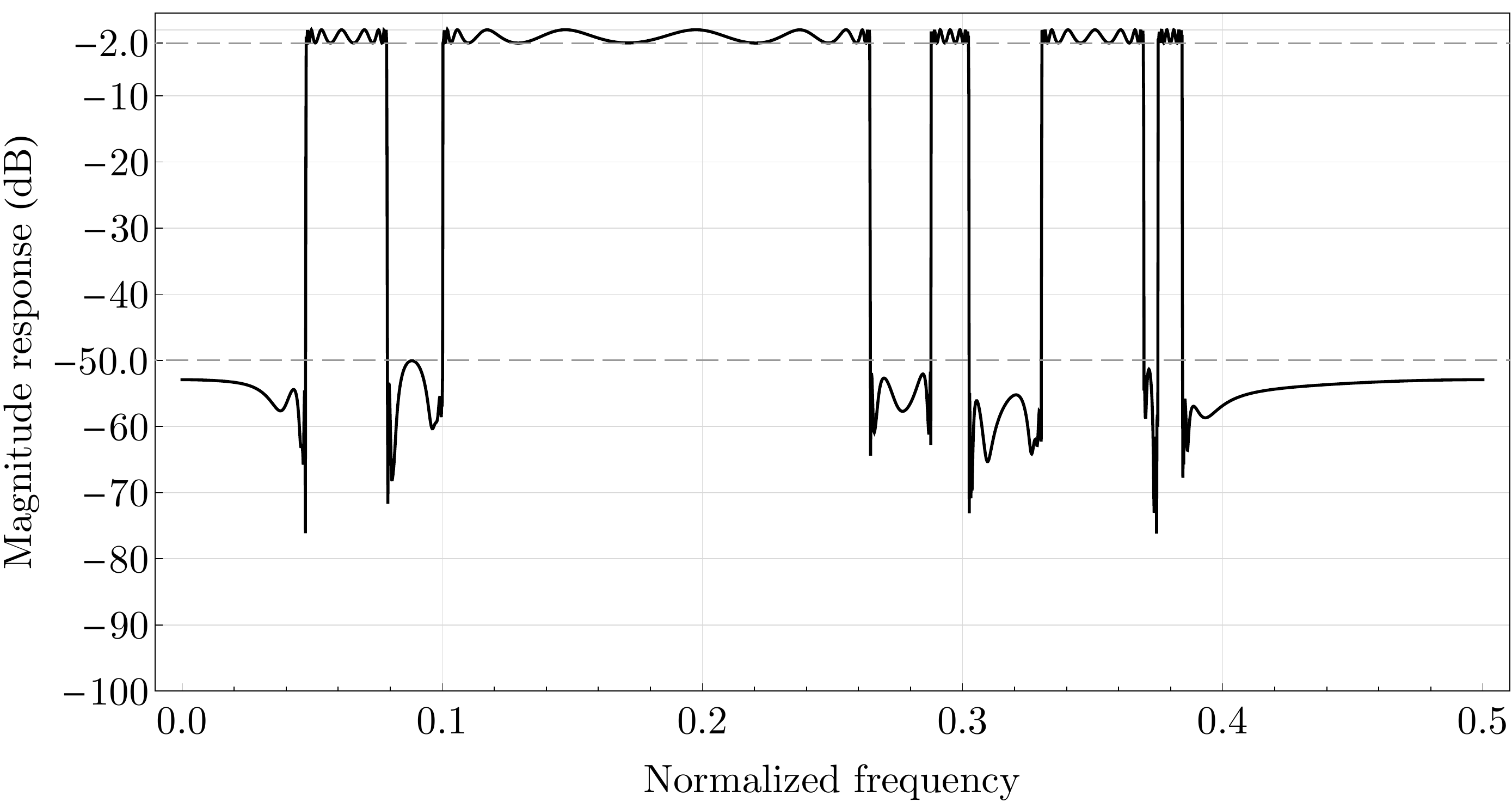}\caption{АЧХ четырёхполосного составного фильтра}\label{5be}}\end{figure}

		Построенный по той же спецификации составной фильтр будет имеет 121. Его АЧХ приводится на Рис. \ref{5be}
	
	\subsection{Фильтр режекторного типа с двумя полосами задержки}	
\begin{figure}[H]{\plot{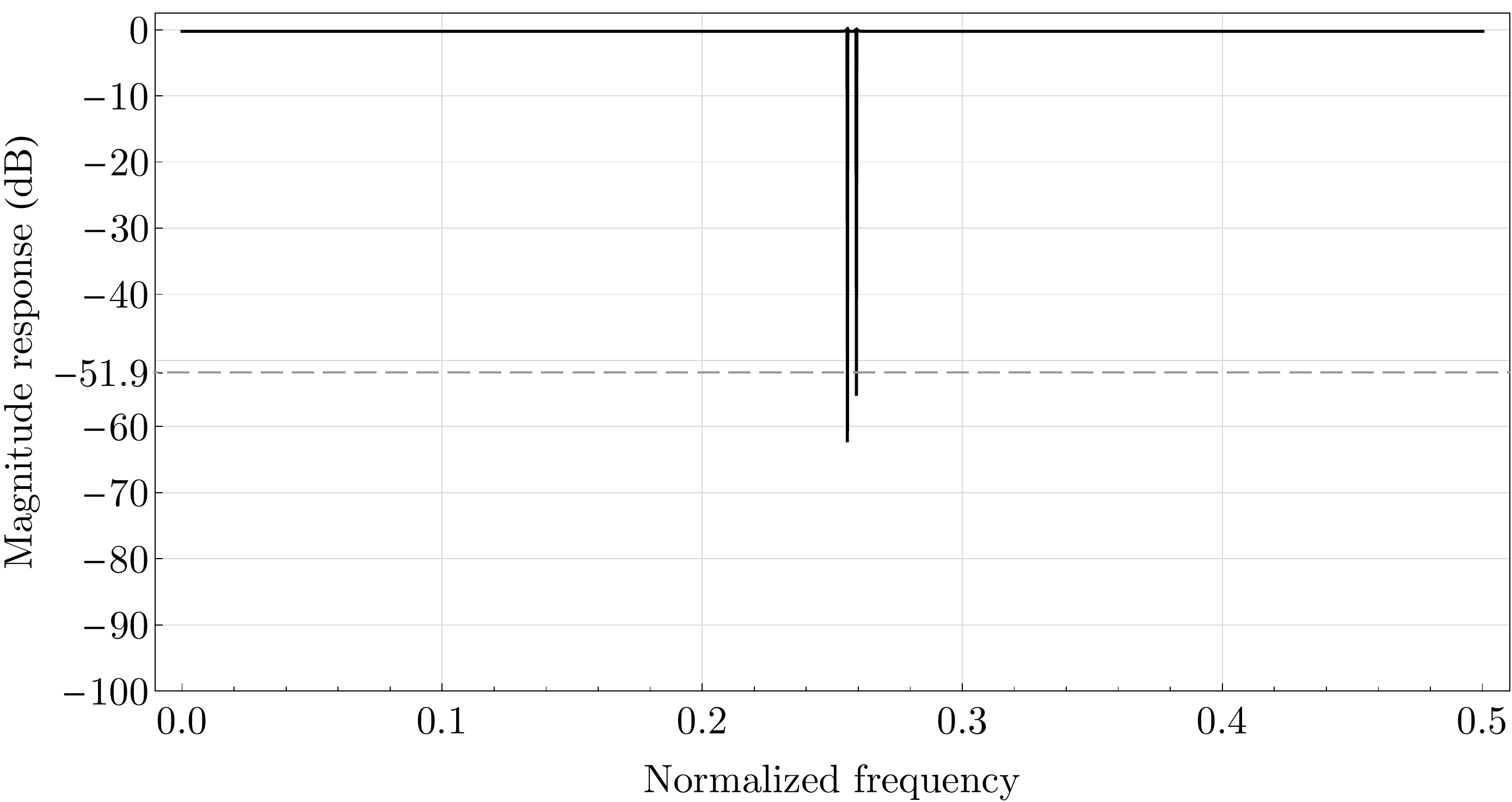}\caption{АЧХ оптимального фильтра режекторного типа с двумя полосами задержки.}\label{2p}}\end{figure}

\begin{figure}[H]{\plot{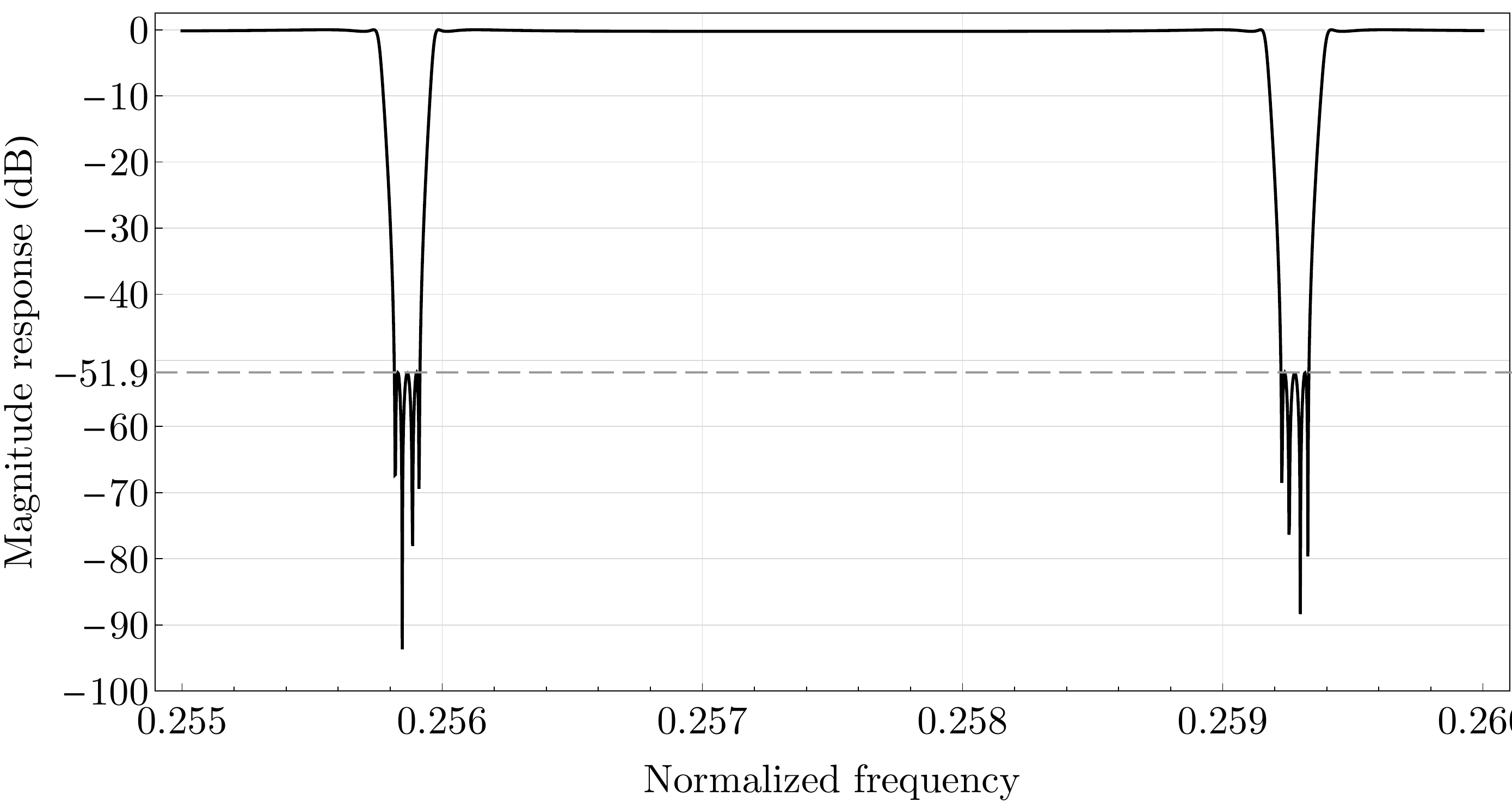}\caption{Участок АЧХ, содержащий вырезаемые частоты.}\label{2p_near}}\end{figure}

		График АЧХ оптимального фильтра 16-ого порядка режекторного типа, осуществляющего точное вырезание двух заданных частот, представлен на Рис. \ref{2p}. Участок АЧХ, содержащий вырезаемые частоты, в большем масшатабе представлен на Рис. \ref{2p_near}.
	
		Составной фильтр, обеспечивающий такое же качество аппроксимации АЧХ, имеет 62-ой порядок.
	
	\subsection{Фильтр с двумя полосами пропускания, расположенными критически близко друг к другу}		
\begin{figure}[H]{\plot{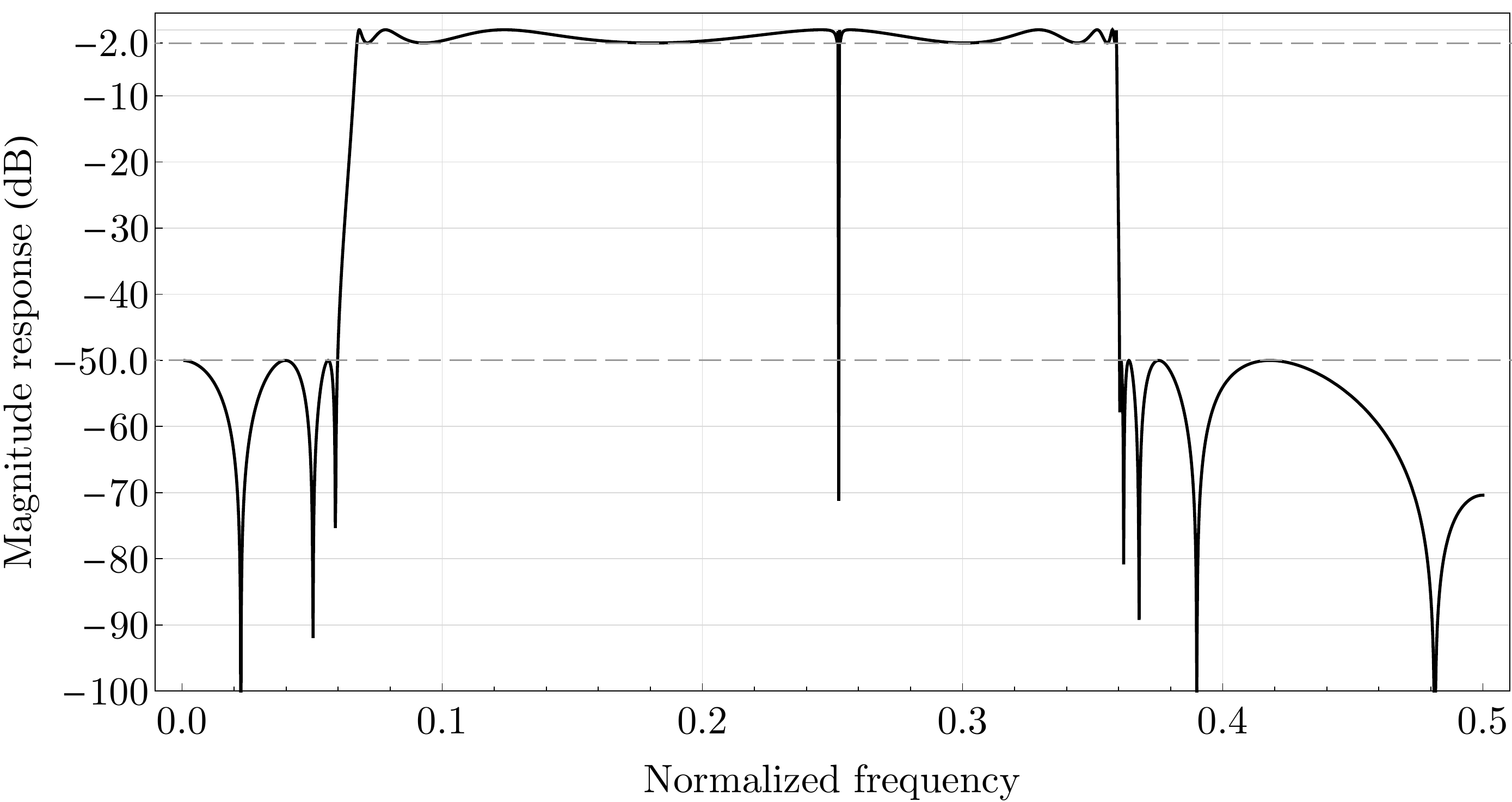}\caption{АЧХ оптимального фильтра с двумя полосами пропускания, расположенными критически близко друг к другу.}\label{1p}}\end{figure}

		С помощью аналитического подхода был рассчитан оптимальный фильтр 24-ого порядка с двумя полосами пропускания, расположенными критически близко друг к другу. Соответствующий график АЧХ приводится на Рис. \ref{1p}. Подобные фильтры могут быть использованы, если требуется выделить из спектра некоторую частотную полосу и при этом задержать одну или несколько частот, сорежащихся внутри этой полосы.
	
		\begin{figure}[H]{\plot{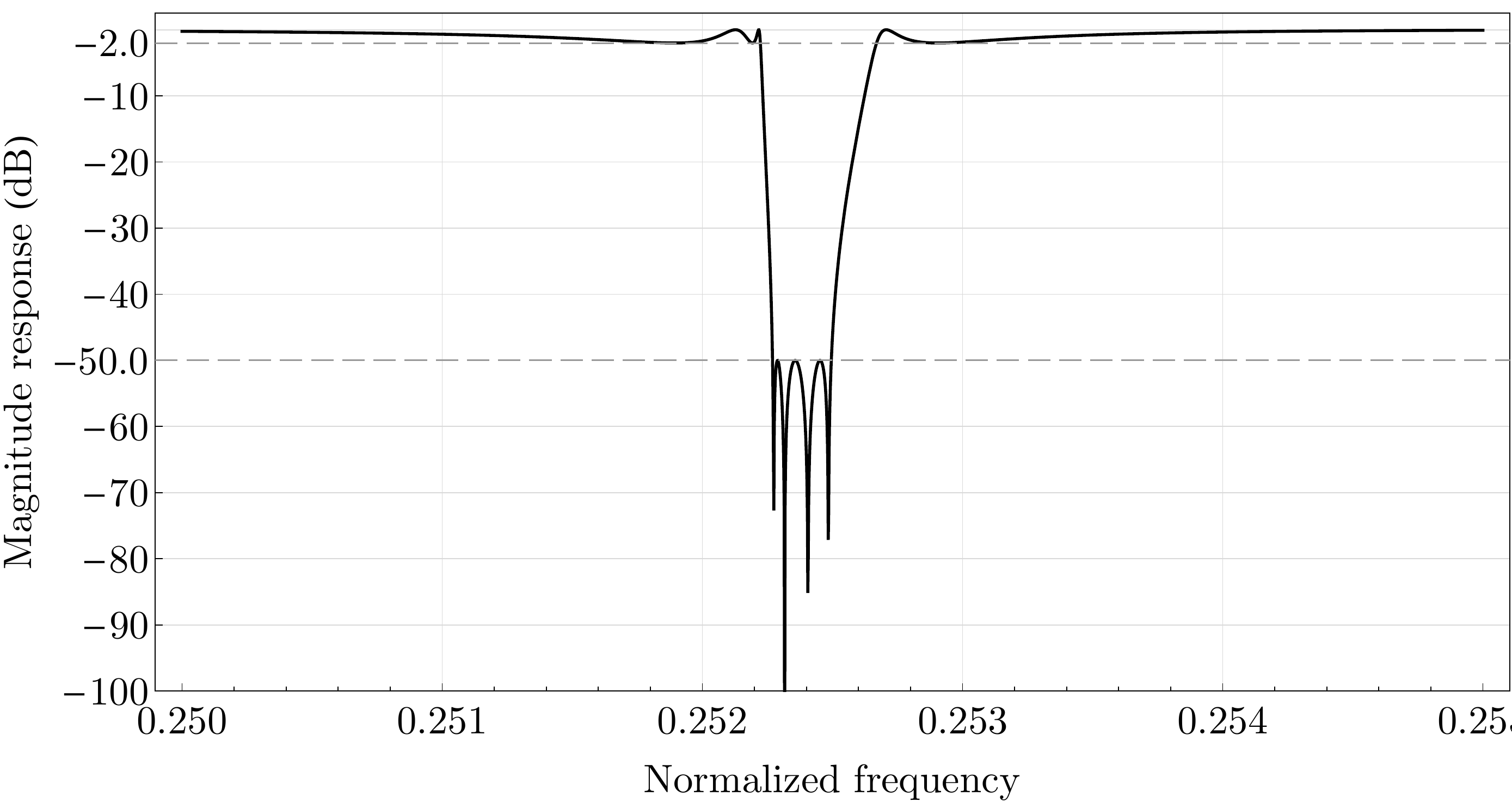}\caption{Участок АЧХ, содержащий вырезаемую частоту.}\label{1p_near}}\end{figure}
	
		Составной фильтр, обеспечивающий такое же качество аппроксимации АЧХ, имеет 59-ый порядок.\\
	
\section{Заключение}

В статье приведено сравнение трех подходов к синтезу многополосных фильтров: нового аналитического подхода, прямой численной оптимизации на основе метода Ремеза и полуаналитический модульный подход.
Прямая численная оптимизация сегодня наиболее алгоритмически проработана: существуют и совершенствуются готовые пакеты для
инженерных расчетов. К сожалению, неустранимая неустойчивость алгоритмов ремезовского типа не позволяет решать слишком сложные
задачи: при использовании двойной точности (15 десятичных знаков) порядок
фильтра не доходит до 20, и при этом невозможно достичь хороших аппроксимационных свойств в случае сложных спецификаций, например, при большом числе полос пропускания и задержки, узких переходных полосах, полосах пропускания, расположенных критически близко друг к другу. Модульный подход состоит в разбиении сложной задачи на ряд простых и их
последовательном решении с использованием дроби Золотарева для
построения АЧХ полосового фильтра. Его преимущество состоит в том, что
таким образом можно гарантированно получить (эрзац-) решение с заданной
спецификацией. Как правило, оно далеко от оптимального: порядок
составного фильтра может в несколько раз превышать порядок
оптимального с той же спецификацией. Ситуация усугубляется с ростом сложности
спецификации фильтра. Наиболее перспективным, с нашей точки зрения, как и
наименее изученным с алгоритмической точки зрения, является
аналитический подход, опирающийся на сложный математический аппарат.
Авторы намереваются продолжать исследования в этом направлении.

\section{Приложение}
\begin{figure}[H]{\plot{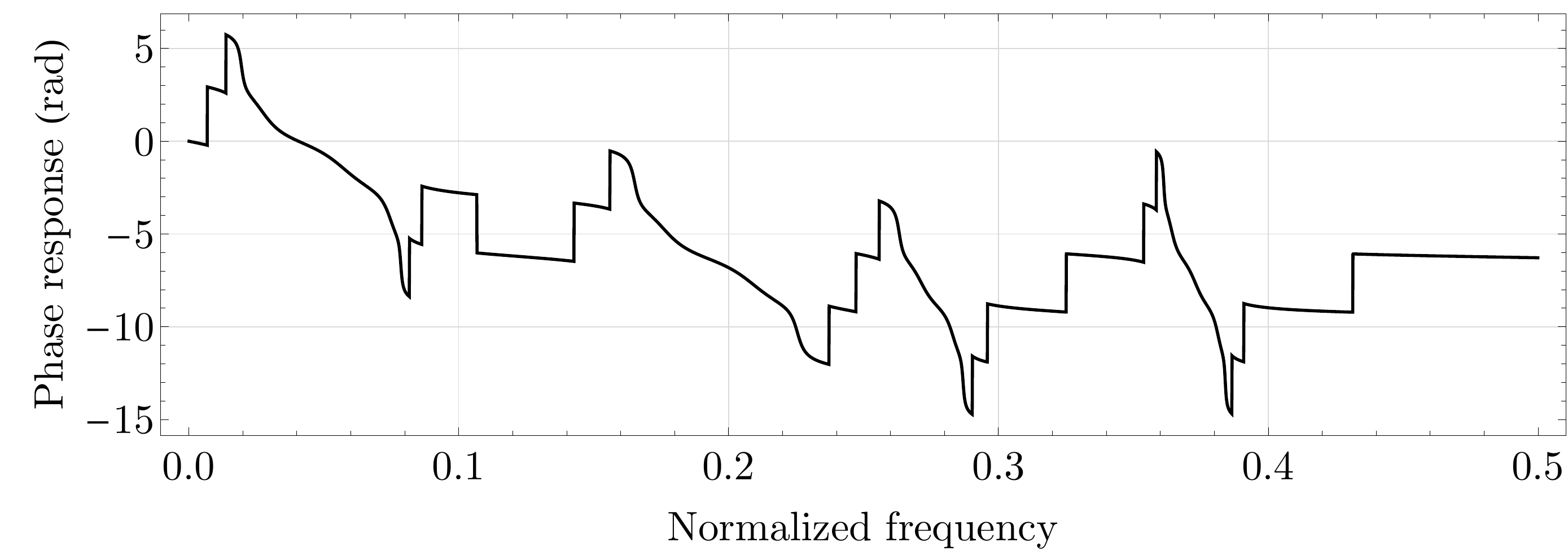}\caption{Фазо-частотная характеристика четырёхполосного оптимального фильтра.}\label{phase}}\end{figure}

\begin{figure}[H]{\plot{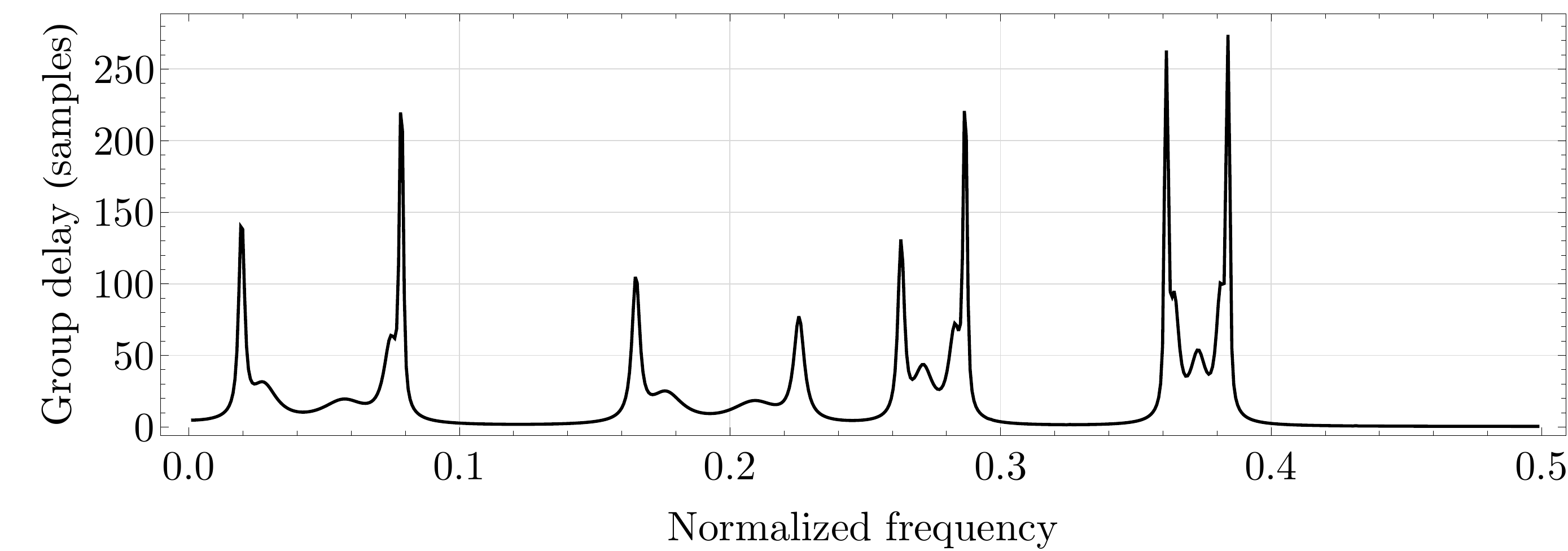}\caption{Групповое время задержки четырёхполосного оптимального фильтра.}\label{gvz}}\end{figure}

\begin{figure}[H]{\plot{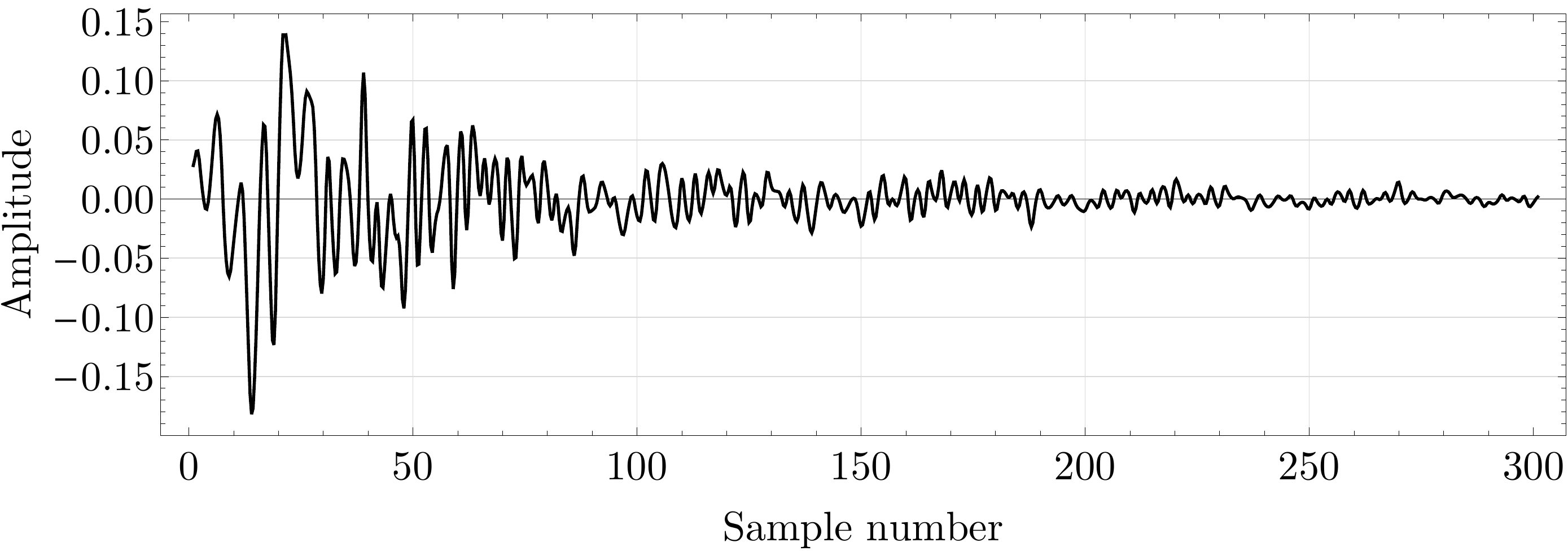}\caption{Импульсная характеристика четырёхполосного оптимального фильтра (точки графика соединены с квадратичной интерполяцией).}\label{ir}}\end{figure}

	На Рис. \ref{phase}, \ref{gvz}, \ref{ir} приводятся графики фазо-частотной и импульсной характеристики, а также группового времени задержки оптимального четырёхполосного фильтра из примера \ref{4bsection}. Его амплитудно-частотная характеристика представлена на Рис. \ref{4b}. 

\vspace{5mm}
\parbox{9cm}
{\it
Институт вычислительной математики РАН,\\
Москва, 119991, ул. Губкина, 8\\[2mm]
{\tt ab.bogatyrev@gmail.com,\\ sergei.goreinov@gmail.com,\\ lyamaev.sergei@gmail.com}}


\begin{thebibliography}{100}
	\bibitem{1} A.B. Bogatyrev, \textit{Chebyshev representation for rational functions}, Sb. Math., 201:11 (2010), pp. 1579–1598.
	\bibitem{2} A. Mohan, S. Singh, A. Biswas, \textit{Generalized synthesis and design of symmetrical multiple passband filters}, Progress In Electromagnetics Research B, Vol. 42, pp. 115-139, 2012.
	\bibitem{11} J. Lee, and K. Sarabandi, \textit{Design of triple-passband microwave filters using frequency transformation}, IEEE Trans. On Microwave Theory and Tech., vol. MTT-56, No. 1, pp 187-193, Jan. 2008. 
	\bibitem{5} V. Lunot, S. Bila, and F. Seyfert, \textit{Optimal synthesis for multi-band microwave filters}, IEEE MTT-S Int. Microw. Symp. Dig., pp. 115–118, Jun. 2007.
	\bibitem{7} D. Deslandes and F. Boone, \textit{An iterative design procedure for the synthesis of generalized dual-bandpass filters},  Int.J. RF Microw. Computer-Aided Eng., vol. 19, no. 5, pp. 607-614, 2009.
	\bibitem{8} G. Macchiarella, \textit{"Equi-ripple"\, Synthesis of Multiband Prototype Filters Using a Remez-Like Algorithm}, IEEE Microwave And Wireless Components Letters, Vol. 23, No. 5, pp. 231-233, May 2013.
	\bibitem{9} R.A.-R. Амer, \textit{The approximation problem of electrical filters}, Basel-Stuttgart, Birkhauser, 1964.
	\bibitem{10} V. Crnojevic-Bengin, \textit{Advances in Multi-Band Microstrip Filters}, Cambridge University Press, 2015.
	\bibitem{B} A.B. Bogatyrev, \textit{Computations in moduli spaces}, Computational Methods and Function Theory, 7 (2007), No. 2, 309-324.  
	\bibitem{B2} А.Б. Богатырев \textit{Экстремальные многочлены и римановы поверхности}, М., МЦНМО, 2005. (монография).
	A.B. Bogatyrev, \textit{Extremal Polynomials and Riemann Surfaces}, Springer Monographs in Mathematics, ISBN 978-3-642-25633-2 (print).
	\bibitem{B3} А.Б. Богатырев, \textit{Конформное отображение прямоугольных семиугольников}, Математический сборник 203:12 (2012), стр 35-56.
	\bibitem{Gonchar} А.А. Гончар, \textit{О задачах Е.И.Золотарева, связанных с рациональными функциями}, Математический сборник 78(120):4 (1969), стр 640-654.
	\bibitem{Remez} Е.Я. Ремез \textit{Основы численных методов чебышевского приближения}, Киев, Наукова думка, 1969. 
	\bibitem{Veidinger} L. Veidinger, \textit{On the numerical determination of the best
	approximations in the Chebyshev sense}, Numer. Math., Vol. 2, pp. 99-105, 1960.
	\bibitem{Boyd} S. Boyd, L. Vandenberghe, \textit{Convex Optimization}, Cambridge University Press, 2004.
	\bibitem{Fuchs} W. Fuchs, \textit{On Chebyshev approximation on sets
	with several components} // D.A. Brannan and J.G. Clunie, eds.,
	Aspects of Contemporary Complex Analysis, pp. 399-408, Academic Press, 1980.
	\bibitem{Zol} Е.И. Золотарев, \textit{Приложение эллиптических функций к вопросам о функциях, наименее и наиболее отклоняющихся от нуля}, Зап. Росс. Акад Наук, ХХХ стр. 37-41, 54-71, 1877. 
	\bibitem{Cauer} W. Cauer, \textit{Theorie der linearen Wechselstromschaltungen}, Bd. 1. Becker und Erler, Leipzig, 1941; Bd. 2. Akademie, Berlin, 1960.
	\bibitem{Malo}  В.Н. Малоземов, \textit{Задача синтеза многополосного электрического фильтра}, Ж. вычисл. матем. и матем. физ., 19:3 (1979),  601–609.
	\bibitem{Akh} Н.И. Ахиезер, \textit{Об одной задаче Е.И. Золотарева}, Изв. Акад. наук СССР, VII серия, 1929, №10, стр. 919--931. 
\end{thebibliography}
\end{document}